\documentclass[namedreferences]{solarphysics}
\usepackage[hyperref,optionalrh,solaromanenum]{spr-sola-addons} % For Solar Physics 
\usepackage{graphicx}                    % For eps figures, newer & more powerfull
\usepackage{color}                       % For color text: \color command
\usepackage{breakurl}                         % For breaking URLs easily trough lines
\usepackage{subfig}
\usepackage{booktabs}
\renewcommand{\vec}[1]{{\mathbfit #1}}

\begin{document}

\begin{article}

\begin{opening}

\title{On the Interaction of Galactic Cosmic Rays with Heliospheric Shocks during Forbush Decreases}

%%%%%%%%%%%%%%%%%%%%%%%%%%%%%%%%%%%%%%%%%%%%%%%%%%%
%% Authors Names
%
 \author[addressref={1},corref,email={akirin@vuka.hr}]{\inits{A.}\fnm{Anamarija}~\lnm{Kirin}\orcid{0000-0002-6873-5076}}
 \author[addressref={2},corref,email={bvrsnak@geof.hr}]{\inits{B.}\fnm{Bojan}~\lnm{Vr{\v s}nak}\orcid{0000-0002-0248-4681}}
 \author[addressref={2}]{\inits{M.}\fnm{Mateja}~\lnm{Dumbovi{\'c}}\orcid{0000-0002-8680-8267}}
 \author[addressref={3}]{\inits{B.}\fnm{Bernd}~\lnm{Heber}\orcid{0000-0003-0960-5658}}

%%%%%%%%%%%%%%%%%%%%%%%%%%%%%%%%%%%%%%%%%%%%%%%%%%%
%% Runningheads
%
\runningauthor{A. Kirin \textit{et al.}}
\runningtitle{On the Interaction of GCRs with Heliospheric Shocks during FDs}

%%%%%%%%%%%%%%%%%%%%%%%%%%%%%%%%%%%%%%%%%%%%%%%%%%%
%% Affilations 
%% id shold be the same with \author addressref value.
\address[id={1}]{Karlovac University of Applied Sciences, Trg J.J.Strossmayera 9, 47000 Karlovac, Croatia}
\address[id={2}]{Hvar Observatory, Faculty of Geodesy, University of Zagreb, Ka{\v c}i{\'c}eva 26,10000 Zagreb, Croatia}
\address[id={3}]{Institut für Experimentelle und Angewandte Physik, Christian-Albrechts-Universit{\"a}t zu Kiel, Christian-Albrechts-Platz 4, 24118 Kiel, Germany}

%%%%%%%%%%%%%%%%%%%%%%%%%%%%%%%%%%%%%%%%%%%%%%%%%%%
%%% Abstract 
\begin{abstract}
Forbush decreases (FDs) are depletions in the galactic cosmic ray (GCR) count rate that last typically for about a week and can be caused by coronal mass ejections (CMEs) or corotating interacting regions (CIRs). Fast CMEs that drive shocks cause large FDs that often show a two-step decrease where the first step is attributed to the shock/sheath region, while the second step is attributed to the closed magnetic structure. Since the difference in size of shock and sheath region is significant, and since there are observed effects that can be related to shocks and not necessarily to the sheath region we expect that the physical mechanisms governing the interaction with GCRs in these two regions are different. We therefore aim to analyse interaction of GCRs with heliospheric shocks only. We approximate the shock by a structure where the magnetic field linearly changes with position within this structure. We assume protons of different energy, different pitch angle and different incoming direction. We also vary the shock parameters such as the magnetic field strength and orientation, as well as the shock thickness. The results demonstrate that protons with higher energies are less likely to be reflected. Also, thicker shocks and shocks with stronger field  reflect protons more efficiently. 
\end{abstract}

%%%%%%%%%%%%%%%%%%%%%%%%%%%%%%%%%%%%%%%%%%%%%%%%%%%
%% Keywords
%

\end{opening}
%-------------------------------------------------

\section{Introduction}
Forbush decreases (FDs) are depletions in the galactic cosmic ray (GCR) count rate that last typically for about a week, which were first reported by \citet{forbush37} and \citet{hess37}. There are two basic types of FDs depending on their cause. Corotating interacting regions (CIRs) cause ``recurrent decreases'' which are lower in amplitude, have a gradual onset and symmetric profile \citep[\textit{e.g.}][]{richardson04}.  Events related to interplanetary coronal mass ejections (ICMEs) cause ``non-recurrent decreases'', which have a sudden onset and a gradual recovery (\textit{e.g.} \citeauthor{lockwood71}, \citeyear{lockwood71}; \citeauthor{cane00}, \citeyear{cane00}; \citeauthor{belov09}, \citeyear{belov09}). For a more detailed comparison of characteristics of the CIR- and ICME- related FDs see \textit{e.g.} \citeauthor{dumbovic12} (\citeyear{dumbovic12}), \citeauthor{badruddin16} (\citeyear{badruddin16}), \citeauthor{melkumyan19} (\citeyear{melkumyan19}).

Fast CMEs that drive shocks cause large FDs with amplitudes up to 30 \% as measured by mid-latitude ground-based neutron monitors (typically observing $\approx$ 1~GeV particles) that often show a two-step decrease. Traditionally, these are regarded as textbook examples of FDs, where the first step is attributed to the shock and a turbulent region which follows it, while the second step is attributed to the closed magnetic structure (\citeauthor{cane00}, \citeyear{cane00}). The observations show that statistically both regions contribute roughly equally to the total FD magnitude, indicating that shock-related FDs are largest due to the cumulative effect of the magnetic structure and disturbed medium in front of it (\citeauthor{richardson11}, \citeyear{richardson11}). The modelling efforts so far were mainly focused on modelling FDs as a single physical region, where the depression occurs due to scattering related to strong turbulence in the shock/sheath region \citep{leroux91,wawrzynczak10} or the effects governed by the shock characteristics \citep[\textit{e.g.},][]{parker61,quenby08,krymsky09}. It was pointed out by Cane \citeyear{cane93} and Wibberenz et al. (\citeyear{wibberenz98}) that it is important to carefully separate the shock/sheath region from the magnetic ejecta region because they are physically different. Indeed, models related to the magnetic structure do separate these two regions (\textit{e.g.} \citeauthor{cane95}, \citeyear{cane95}; \citeauthor{munakata06}, \citeyear{munakata06}; \citeauthor{subramanian09}, \citeyear{subramanian09}; \citeauthor{dumbovic18b}, \citeyear{dumbovic18b}). However, in order to fully understand the FD, there are strong indications that the shock itself should be decoupled from the sheath region as well.

The interplanetary shocks at 1~AU are practically a discontinuities with typical size $\approx 10^4 $~km \citep{pinter80}, which is much smaller than the gyro-radius of the primary GCR corresponding to the detection by ground-based neutron monitor (1~GeV proton in 10~nT field $\approx 10^6 $~km). The sheath region at 1~AU is however much larger in size (typically $\approx 10^7 $~km at Earth based on \citeauthor{russell02}, \citeyear{russell02}) and comparable to the gyro-radius of 1~GeV protons. Therefore, we might expect that the physical mechanisms governing the interaction with GCRs in these two regions are different. The two-step FDs observed in neutron monitors very often show a precursory increase, possibly resulting from reflection of particles from the shock or acceleration at the shock, although the latter is not considered likely even for very strong shocks (\citeauthor{cane00}, \citeyear{cane00}). Very strong shocks are often associated with series of pre-decreases and pre-increases in the CR intensity accompanied by the changes in the first harmonic of the anisotropy at the ecliptic plane appearing hours before the FD onset (see \textit{e.g.} \citeauthor{belov95}, \citeyear{belov95}, \citeauthor{papailiou12}, \citeyear{papailiou12}, \citeauthor{lingri19}, \citeyear{lingri19}, and references therein). From that perspective, since also GCR increase can be observed along with FD, a term Forbush effect might be more proper \citep{belov09}. The recovery phase of FDs lasts long after the ICME has passed, can frequently be well described as exponential \citep{lockwood86} and can be explained through the global effect of the shock, so called ``shadow effect'', where the shock front casts a smaller and smaller ``shadow'' on the observer as it moves away (for more details see \citeauthor{lockwood86}, \citeyear{lockwood86}; \citeauthor{dumbovic11}, \citeyear{dumbovic11}). These observations are likely to be related to the global geometry of the interplanetary shock, and not necessarily related to the sheath region that follows.

Based on these considerations, we hypothesize that the shock region should be decoupled from the sheath region when analysing FDs and therefore aim to analyse interaction of GCRs with heliospheric shocks in the context of the FD observations noted above. For that purpose we limit the analysis to high-energy protons ($\approx$ 1~GeV) which typically produce secondary neutrons detected by ground-based neutron monitors. The heliospheric shock structure is quite complex and each of its segments - a foot, a ramp, and an overshoot region (\textit{e.g.} \citeauthor{stone85}, \citeyear{stone85}) might influence the GCR, however, we will focus on a ramp as the most pronounced feature of the shock. We focus on a single GCR proton, and adopt a test particle approach. We do not adopt a full-orbit particle simulations usually used in particle-in-cell (PIC) models to simulate particle acceleration (\textit{e.g.} \citeauthor{giacalone04}, \citeyear{giacalone04}) since at 1~AU the collective properties arising from the interaction of the shock at relevant energies ($\approx$ 1~GeV) can be regarded as negligible, as can the particle acceleration at these energies. Acceleration to the GeV energies can occur near the Sun, where the background magnetic field is large, and the intensity of self-excited waves is largest. On the other hand, at 1~AU the background field is much weaker, the wave growth is not as rapid, and the waves are less intense for 1~GeV particles (\citeauthor{desai16}, \citeyear{desai16}). We next assume that the shock structure is laminar and smooth, since any fluctuations can be maximally of the shock size ($\approx 10^4 $~km), \textit{i.e.} much smaller than the gyro-radius of our typical particle ($\approx 10^6 $~km) and therefore our particle will not see it. We strongly highlight that the assumptions used here are suitable only for the context of highly energetic protons ($\approx$ 1~GeV) interacting with interplanetary shocks typically observed at 1 AU and are therefore not suitable for considerations of \textit{e.g.} solar energetic particles, for which the modelling assumptions and approach should be much different (see \textit{e.g.} \citeauthor{desai16}, \citeyear{desai16}).

It is our intention to analyse GCR-shock interaction depending on key shock characteristics (shock thickness, upstream field strength, $B_{1}$, downstream to upstream field strength ratio, $B_{2}/B_{1}$, and an inclination angle $\tan \theta =B_{1y}/B_{1x} $) and particle properties (energy, initial speed components) in order to better understand the shock-related Forbush effect. While consequently our analysis might lead to a theoretical model describing the shock-related Forbush effect we note that building such a model would include substantial integration over all incoming directions, for protons of various energies, which is beyond the scope of this paper and will be given elsewhere.

\section{Model}
Observationally, ICME-related interplanetary shocks resemble fast-mode MHD shocks, since the downstream magnetic field is stronger than the upstream (\textit{e.g.} \citeauthor{kilpua17}, \citeyear{kilpua17}). Therefore, we consider an oblique 2-dimensional (2D) MHD fast-mode shocks on the  GCR protons, where the magnetic field contains components both parallel and normal to the shock front (Figure~\ref{fig1}). Regarding the modeling of the high-energy protons behaviour, in such a situation there are several issues that have to be emphasized:

\begin{figure}[h!]
\centering
\subfloat[]{\includegraphics[width=0.35\textwidth]{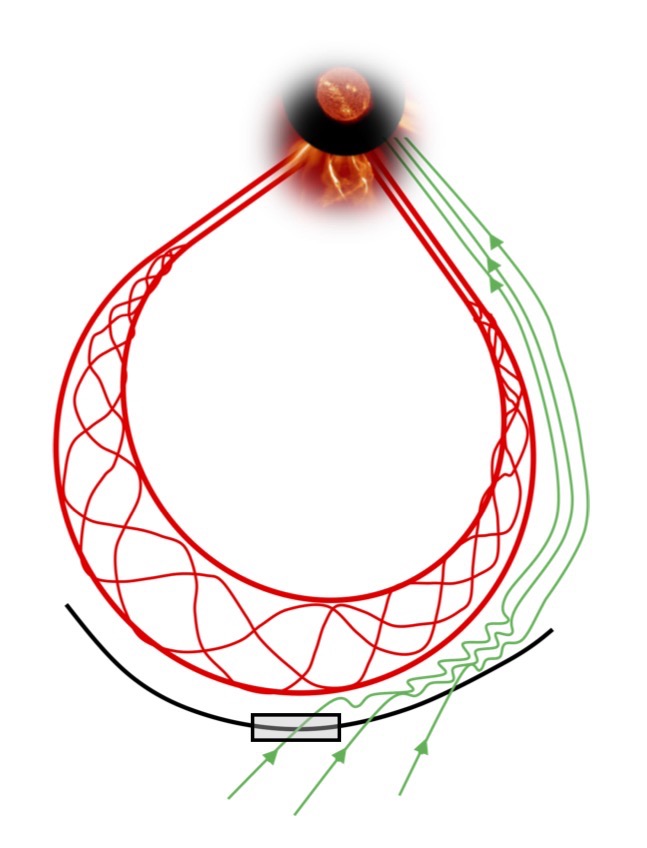}\label{fig1a}}
\subfloat[]{\includegraphics[width=0.65\textwidth]{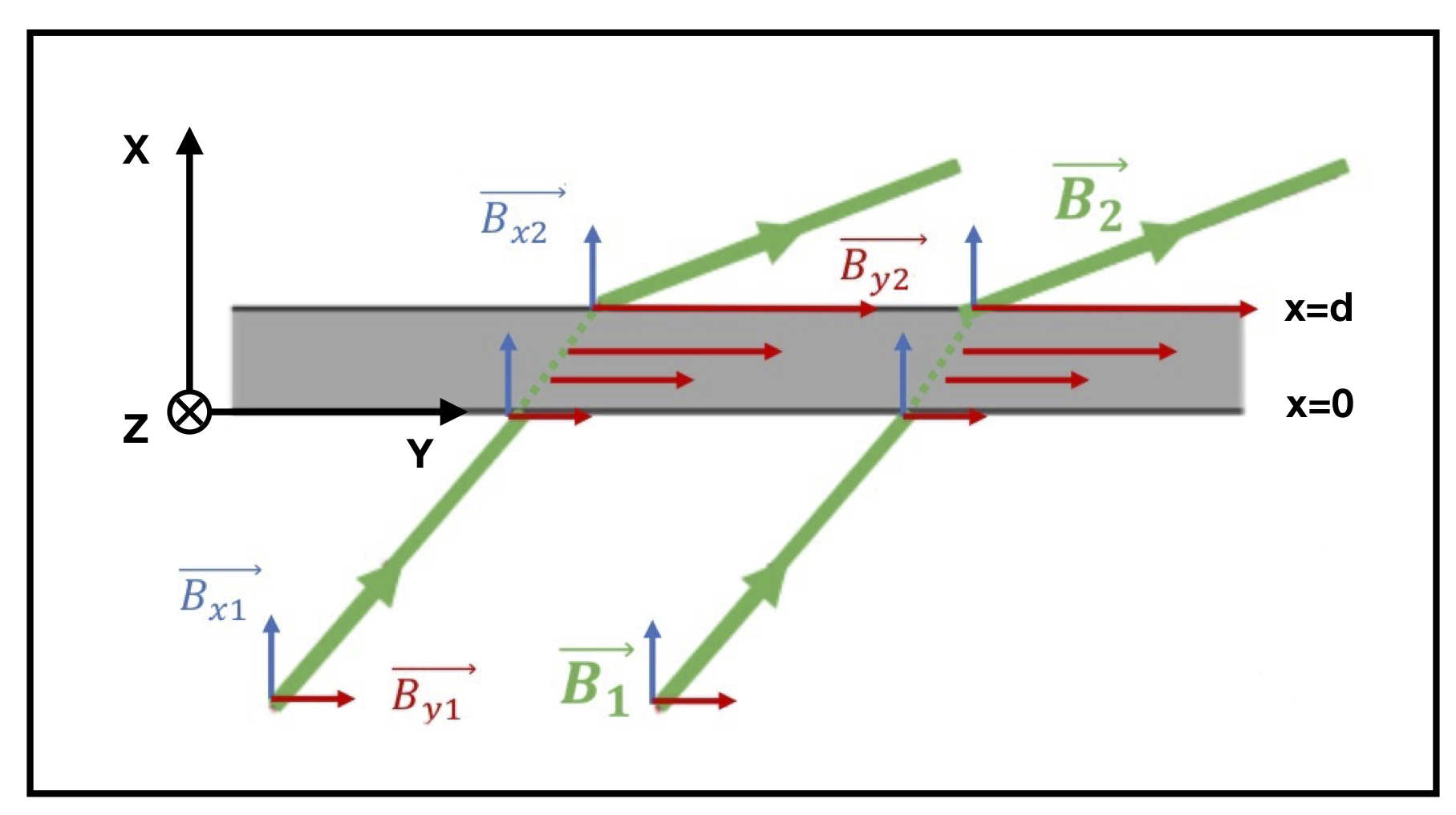}\label{fig1b}}\\
\caption{a) Schematic of an interplanetary coronal mass ejection driving a shock ahead of it; b) Enlarged shock region: change of the magnetic field configuration inside an oblique fast-mode MHD shock. Magnetic field in the upstream region ($x<0$) is weaker than magnetic field in the downstream region ($x>d$).}
\label{fig1}
\end{figure}

\begin{figure}[h!]
\centering
\includegraphics[width=0.8\textwidth]{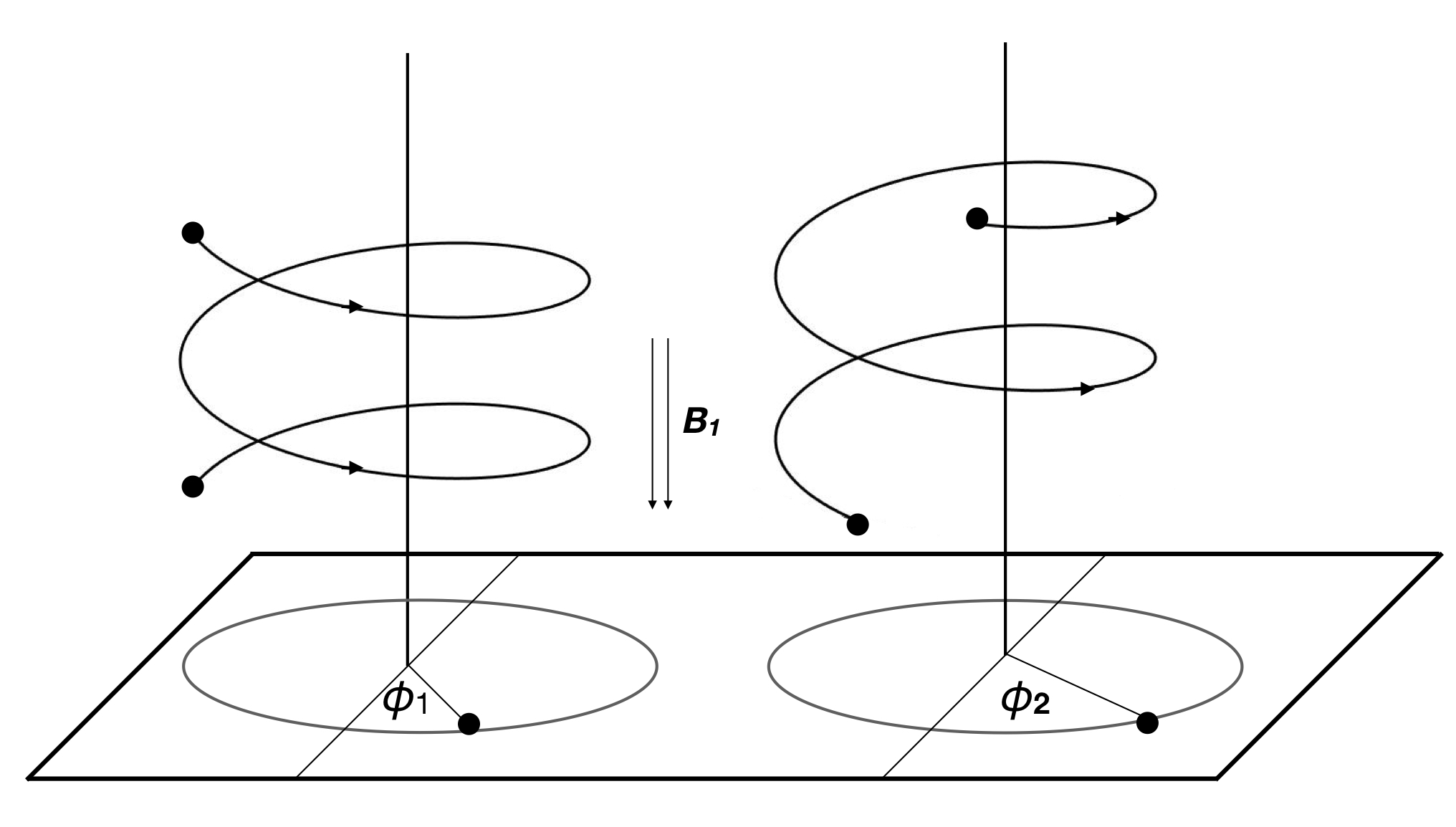}
\caption{Azimuthal angle $\phi$. Particles with different azimuthal angles have different values of speed components at the moment they enter the shock and therefore behave differently inside the shock.}
\label{fig2}
\end{figure}

\begin{itemize}
\item Since the shock profile is extremely sharp, \textit{i.e.} its thickness is much smaller than the proton gyro-radius, the ``standard'' magnetic mirror approximations cannot be applied;
\item Due to this, the behaviour of protons turns out to be dependent not only on the upstream/downstream magnetic strength ratio and particle pitch angle, but also on the shock thickness (unlike in the case of the ``standard'' magnetic mirror approximation);
\item We assume an isotropic GCR flux, therefore we allow the test particle to have an arbitrary direction with respect to the shock front;
\item The direction of the particle with respect to the shock front does not depend only on the pitch angle, but also on its energy and the impact azimuthal angle $\phi$, (see Figure~\ref{fig2}) \textit{i.e.}, it depends on the impact direction relative to the magnetic field and the shock surface;
\item Consequently, since particle pitch angle alone does not define the direction of the incoming particle with respect to the shock front, for practical reasons, the simplest option for describing the proton behaviour is to follow its trajectory in the Cartesian coordinates defined by the shock characteristic directions. 
\end{itemize}

The coordinate system is set so that \textit{x}-axis is parallel to the shock normal and \textit{y}-axis is perpendicular to it. According to the magnetic flux conservation the \textit{x}-component of magnetic field doesn't change. We take the \textit{z}-component of magnetic field to be zero. The y-component changes from value $B_{1y}$ ahead of the shock (upstream region) to $B_{2y}$ behind the shock (downstream region). Since the magnetic field of fast-mode MHD shock in the downstream region is stronger than magnetic field in the upstream region and the \textit{x}-component doesn't change, it must be $B_{2y}>B_{1y}$. In our model, we consider the simplest case where the change in y-component of the field across the shock is linear, \textit{i.e.}
\begin{equation}
B_{y}(x)=B_{1y}+(B_{2y}-B_{1y})\frac{x}{d},\quad 0 \leq x \leq d.
\end{equation}
Here $B_{y}(x)$ is the \textit{y}-component of the magnetic field inside the shock, $d$ is the shock thickness, and $x$ is the distance from the shock upstream border so that $B_{y}(0)=B_{1y}$ and $B_{y}(d)=B_{2y}$. 

In the upstream and downstream region the field is homogeneous, so the proton moves in a helical path along the field line. However, the field inside the shock is not homogeneous, so the helical motion of the proton is disrupted (the total speed remains the same since the stationary magnetic field cannot change the proton energy). 

The equation of motion 
\begin{equation}
\vec{F}=m\vec{a}=q\vec{v}\times\vec{B}
\end{equation}
written in components inside the shock implies:
\begin{equation}
a_{x}=-\frac{q}{m}v_{z}B_{1y}-\frac{q}{m}v_{z}\frac{\Delta B_{y}}{d}x,
\label{ax}
\end{equation}
\begin{equation}
a_{y}=\frac{q}{m}v_{z}B_{x},
\end{equation}
\begin{equation}
a_{z}=\frac{q}{m}v_{x}B_{1y}-\frac{q}{m}v_{y}B_{x}+\frac{q}{m}v_{x}\frac{\Delta B_{y}}{d}x,
\end{equation}
where $\Delta B_{y}=B_{2y}-B_{1y} $, and $m=m_0 \gamma$  with $\gamma = \sqrt{\frac{1}{1-(\frac{v}{c})^2}}$ the Lorentz factor for a proton with speed $v$ and \textit{c} is the speed of light. 

In the upstream and downstream region components of the equation of motion are respectively
\begin{equation}
a_{x}=-\frac{q}{m}v_{z}B_{1y},
\end{equation}
\begin{equation}
a_{y}=\frac{q}{m}v_{z}B_{x},
\end{equation}
\begin{equation}
a_{z}=\frac{q}{m}v_{x}B_{1y}-\frac{q}{m}v_{y}B_{x},
\end{equation}

and

\begin{equation}
a_{x}=-\frac{q}{m}v_{z}B_{2y},
\end{equation}
\begin{equation}
a_{y}=\frac{q}{m}v_{z}B_{x},
\end{equation}
\begin{equation}
a_{z}=\frac{q}{m}v_{x}B_{2y}-\frac{q}{m}v_{y}B_{x}.
\end{equation}

We use the Ordinary Differential Equation (ODE) solver in MATLAB$^{®}$ to solve these equations and to obtain proton trajectories for different initial speed components (initial in this context means at the upstream shock border), \textit{i.e.} $v_{0x}$, $v_{0y}$ and $v_{0z}$, for protons of different energies and also for shocks with different parameters (shock thickness, upstream field strength, $B_{1}$, downstream to upstream field strength ratio, $B_{2}/B_{1}$, and an inclination angle $\tan \theta =B_{1y}/B_{1x} $). The used integration method is numerically stable and does not introduce numerical energy gain. However, we note that the numerical energy gain would not affect our results, since this effect happens with each gyration, and in our simulation the gyro-radius is much larger than the shock thickness. The initial conditions correspond to initial speed components. Since a proton of a given energy has a certain total speed, two of its components determine the third. Therefore in the rest of the paper, only proton energy and two speed components will be given and the third one can be easily calculated. We choose to work with $v_{0x}$ and $v_{0z} $ because, for proton to become reflected, $v_{0x}$ needs to decrease and turn negative, and the acceleration in \textit{x}-direction contains only $v_{0z}$ and not $v_{0y} $. Trajectories for given shock parameters allow us to see how a certain shock affects protons of different energies, different pitch angles and different incoming directions. In all of the following figures distance in \textit{x}-direction is scaled to shock thickness $ d $, so the shock thickness is always one, and a distance of for example 10 in \textit{x}-direction is actually $ 10 \, d $. Negative values of \textit{x} represent the upstream region, in particular the distance from the shock border at $x=0$.

\section{Test Proton Examples}
\begin{figure*}[h!]
\centering
\subfloat[]{\includegraphics[width=0.5\textwidth]{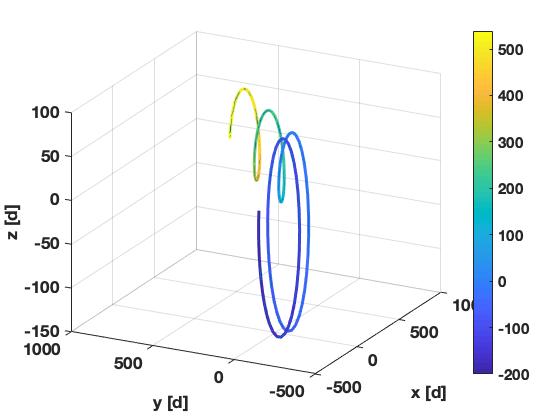}}
\subfloat[]{\includegraphics[width=0.5\textwidth]{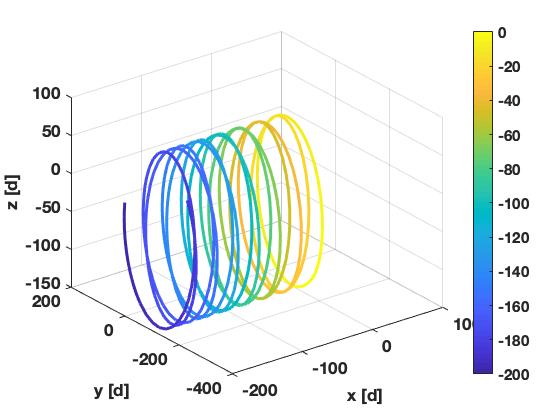}}
\caption{Effect of a longitudinal shock to a 1~GeV proton motion. The initial speed is $v_{0}=0.875 c$ and in both cases $v_{0z}=-250\ 000$~km~s$^{-1}=-0.834 c$.  Shock parameters are: $B_{1}=5$~nT, $B_{2}/B_{1}=1.5$, $d=10\ 000$~km; a)~Proton crosses the shock and continues its motion in downstream region with smaller Larmor radius due to the stronger magnetic field. The \textit{x}-component of the initial speed is: $v_{0x}=40\ 000$~km~s$^{-1}=0.133 c$ and the pitch angle is $\vartheta=81.25 ^\circ $; b)~Proton is reflected inside the shock, and returns to the upstream region. The \textit{x}-component of the initial speed is: $v_{0x}=10\ 000$~km~s$^{-1}=0.033 c$ and the pitch angle is $\vartheta=87.84 ^\circ $.}
\label{fig31}
\end{figure*}

\begin{figure*}[h!]
\centering
\subfloat[]{\includegraphics[width=0.5\textwidth]{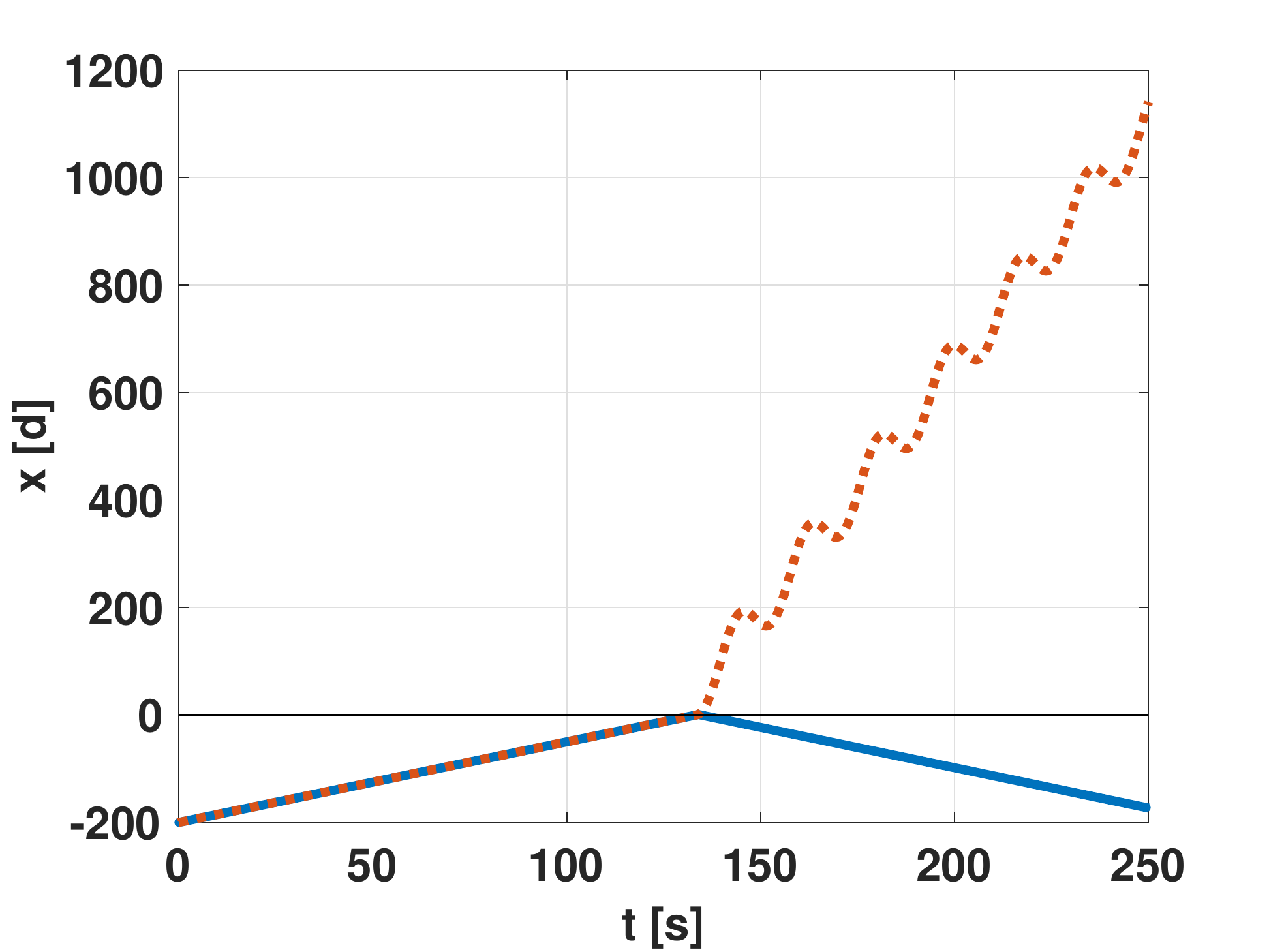}}
\subfloat[]{\includegraphics[width=0.5\textwidth]{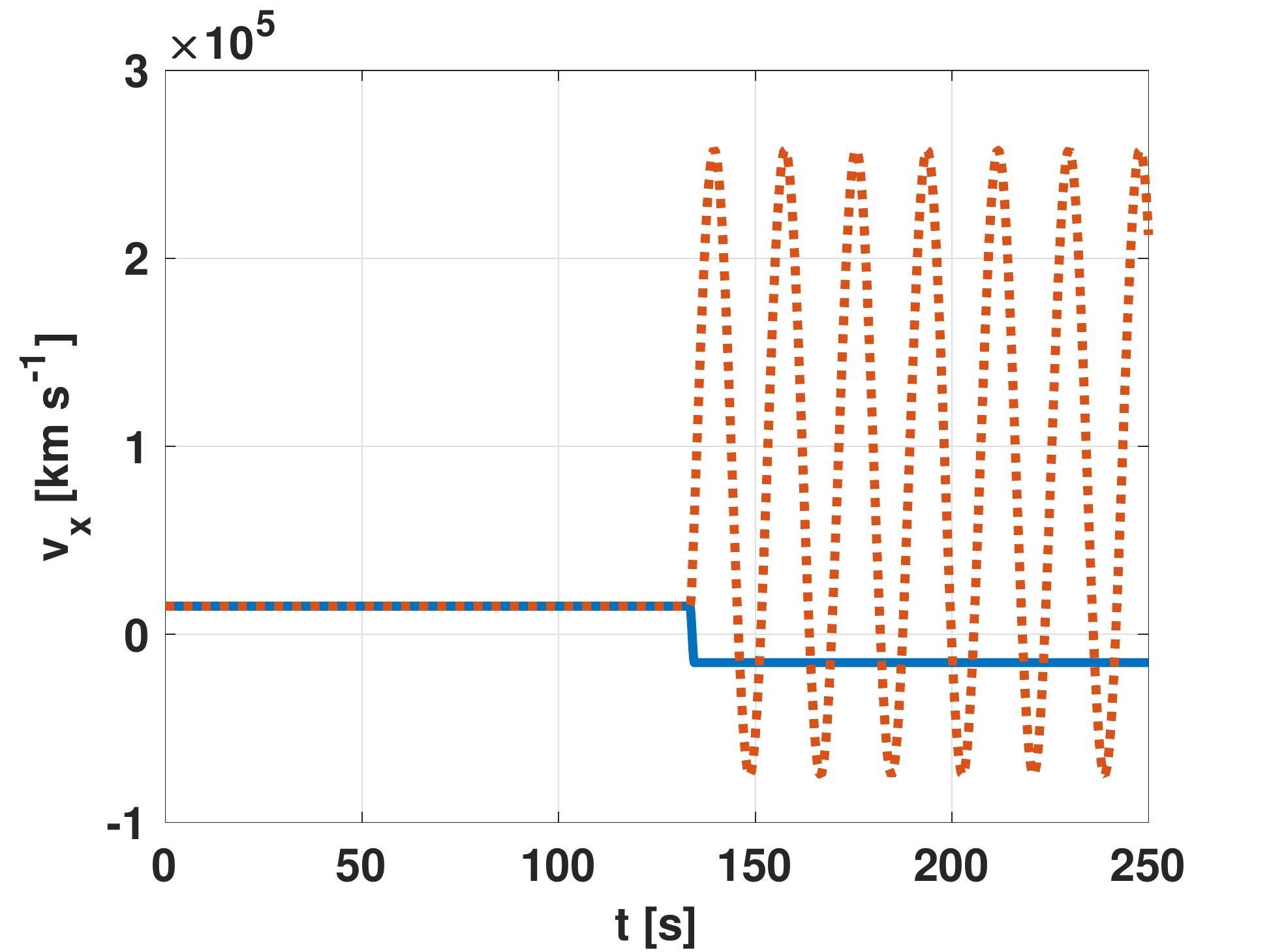}}
\caption{a) Motion in \textit{x}-direction of two protons with different initial values $v_{0z}$. Dashed red line represents transmitted proton ($v_{0z}=-250\ 000$~km~s$^{-1}=-0.834 c$) and blue solid line a reflected proton ($v_{0z}=250\ 000 $~km~s$^{-1}=0.834 c$). b) Change of $v_{x}$ component of proton speed for the same two protons. The initial value of $v_{x}$ is the same in both cases, \textit{i.e.} $v_{0x}=15\ 000$~km~s$^{-1}=0.050 c$ and therefore the pitch angle is $\vartheta=86.72 ^\circ $. Since the shock is a longitudinal shock, the \textit{x}-component of the speed is a constant in the upstream region. Shock parameters are: $B_{1}=5$~nT, $B_{2}/B_{1}=1.5$, $d=10\ 000$~km.}
\label{fig32}
\end{figure*}

\begin{figure*}[h!]
\centering
\subfloat[]{\includegraphics[width=0.5\textwidth]{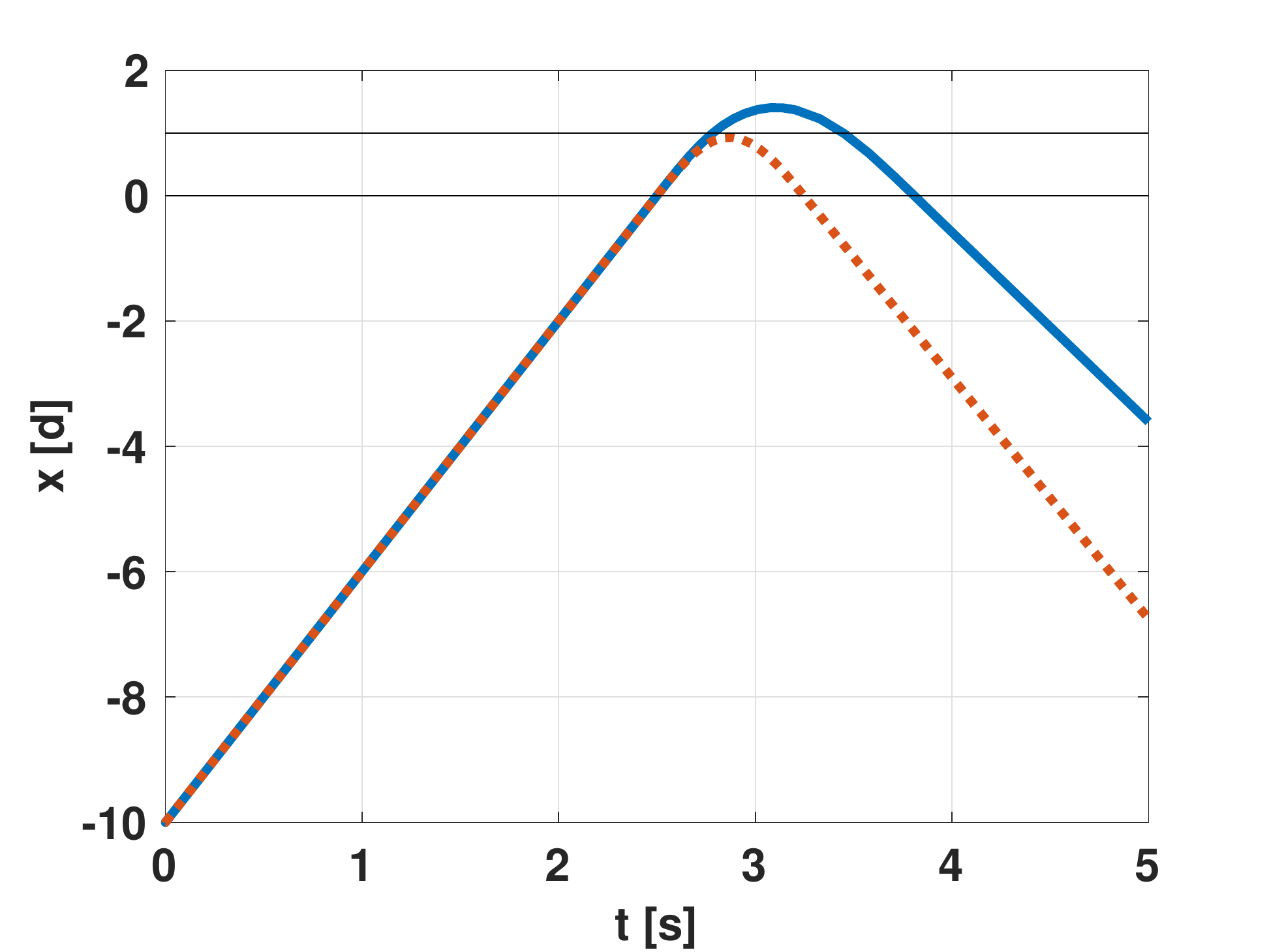}}
\subfloat[]{\includegraphics[width=0.5\textwidth]{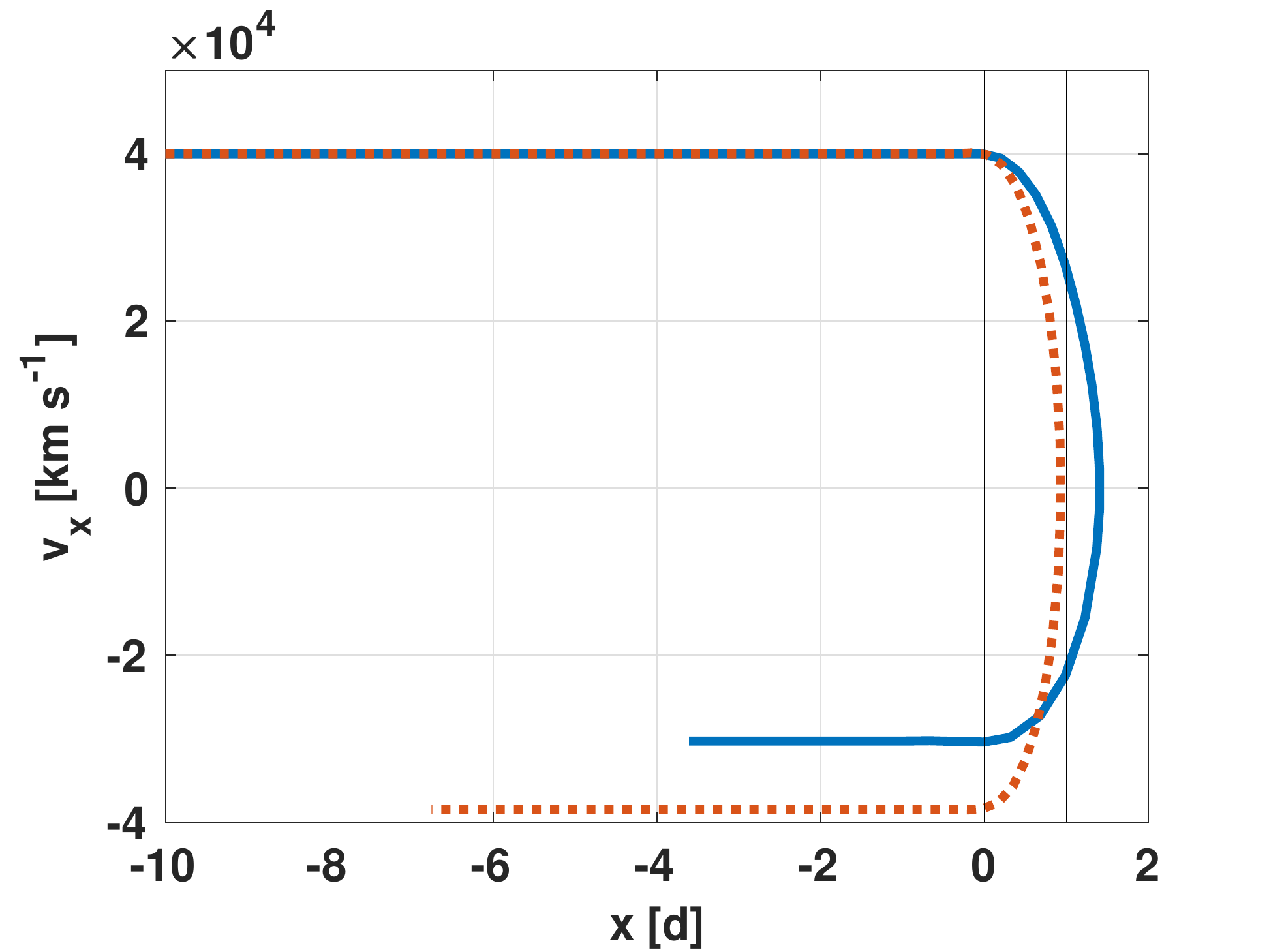}}
\caption{Comparison of a) trajectories and b) \textit{x}-components of the speed, $v_{x}$, for protons reflected inside the longitudinal shock (red dashed line) and behind the same shock in the downstream region (blue solid line). Both protons have the same initial value of $v_{x}$, $v_{0x}=40\ 000$~km~s$^{-1}=0.133 c$ and the pitch angle $\vartheta=81.25 ^\circ $, but for the proton reflected inside the shock initial value of $v_{z}$ is larger \textit{i.e.}  $v_{0z}=259\ 000$~km~s$^{-1}=0.864 c$, and for the proton reflected behind the shock $v_{0z}=220\ 000$~km~s$^{-1}=0.734 c$. Shock parameters are: $B_{1}=5$~nT, $B_{2}/B_{1}=4$, $d=10\ 000$~km.}
\label{fig33}
\end{figure*}

As already mentioned, a GCR proton moves in a helical path along the field lines in a homogeneous magnetic field ahead of the shock. After entering the shock, the field is no more homogeneous and the trajectory of the proton starts to change. In this respect we note that the shock thickness is very small compared to the helix radius. Nevertheless, the effect can be quite significant and easily demonstrated (Figure~\ref{fig31}) if we look at the proton trajectory before and after the interaction with shock, where ``after" doesn't have to mean ``downstream region" because the proton can be reflected back to the upstream region. The simplest case to demonstrate this is that of a longitudinal shock, where the field in the upstream region is parallel to the shock normal, \textit{i.e.} the $B_{y}$ component is zero. In this case there is no acceleration in the \textit{x}-direction, and the $v_{x}$ component of the proton speed doesn't change in the upstream region (Figure~\ref{fig32}). The change starts inside the shock where the proton can be reflected back to upstream region by decreasing the $v_{x}$ to negative values or transmitted to downstream region (Figure~\ref{fig32}). Once the proton is reflected to the upstream region with negative $v_{x}$, it continues to move away from the shock without ever coming back. However, since the field lines in the downstream region are not parallel to the shock front, it is possible for a proton to enter the shock region again, due to its helical motion. Then it can be easily accelerated back to the upstream region (Figure~\ref{fig33}, blue line). We note that the fact that the proton accelerates in the \textit{x}-direction does not necessarily imply energy gain, because it decelerates in the \textit{y}-direction at the same time, and also no electric fields are assumed in these simulations. Alternatively, it is possible for a proton to continue spiral motion in downstream region without ever returning to shock and upstream region (Figure~\ref{fig32}, red dashed line). The proton and field parameters in previous examples have no special meaning, they are simply chosen in such a way that resulting figures clearly demonstrate described behaviours. Their values are given in the figure descriptions.

\begin{figure*}[h!]
\centering
\includegraphics[width=0.8\textwidth]{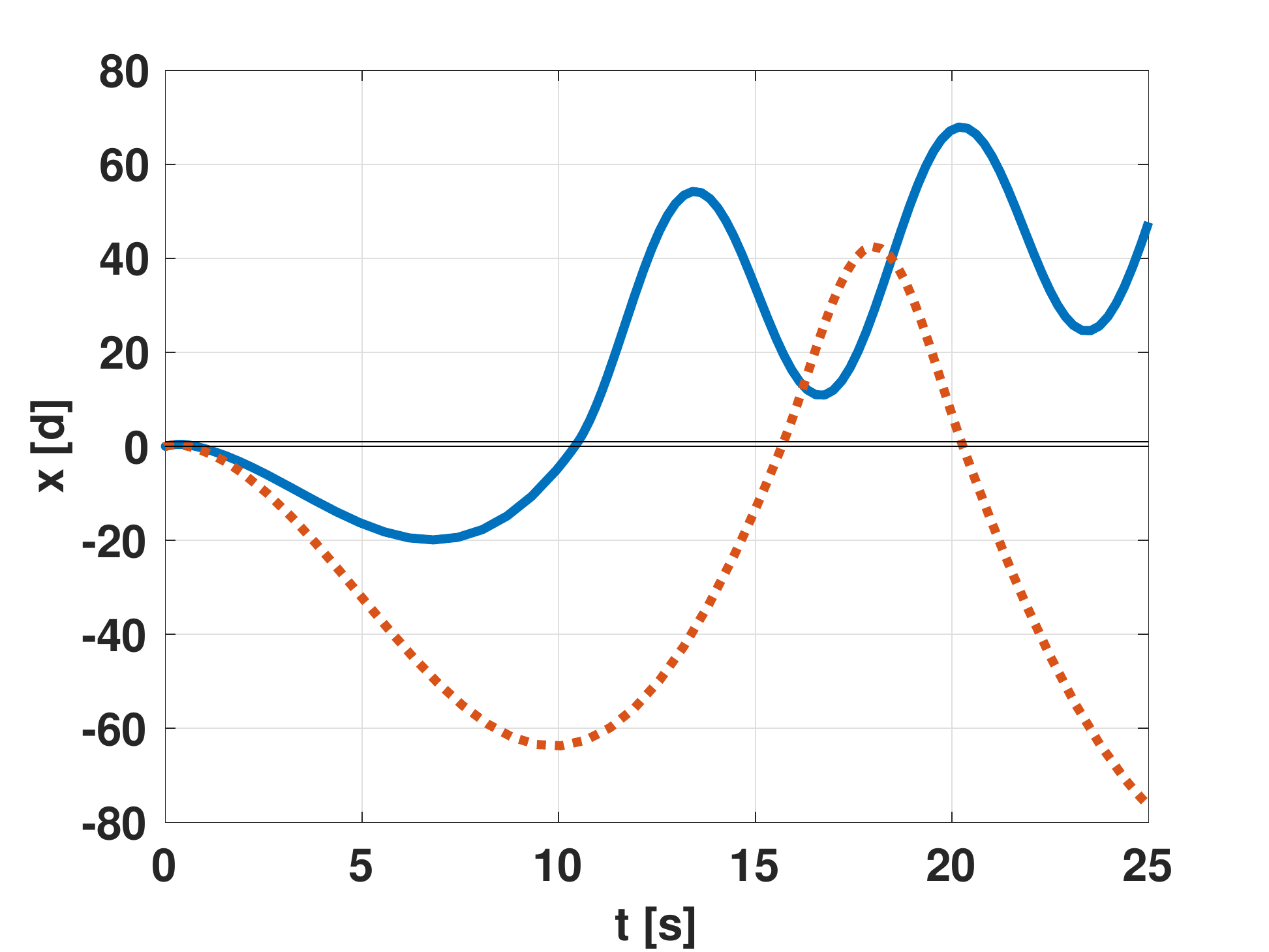}
\caption{1~GeV proton trajectories after being reflected inside the shock of strength $B_{2}/B_{1}=4$, $B_{1}=5$~nT, inclination angle $\theta =45^{\circ} $ and thickness $d=10\ 000$~ km. Both protons ($v_{0x}=20\ 000$~km~s$^{-1}=0.067 c$) are reflected inside the shock back to the upstream region but then enter the shock again. This time they pass the shock and enter the downstream region. One proton continues its motion in downstream region (blue line, $v_{0z}=150\ 000$~km~s$^{-1}=0.500c$, $\vartheta=50.81 ^\circ $) and the other is again reflected to the upstream region where it continues its motion (red line, $v_{0z}=220\ 000$~km~s$^{-1}=0.734 c$, $\vartheta=64.20 ^\circ $). }
\label{fig34}
\end{figure*}

In the general case, for a given field configuration we have similar scenarios. In all of them, a proton is coming from upstream region towards the shock, \textit{i.e.} it has positive \textit{x}-component of initial speed, $v_{0x}>0$. Then we have three options: 
\begin{enumerate}
\item The \textit{x}-component of proton speed drops to zero inside the shock and proton is reflected back to the upstream region with $v_{x}<0$ where it continues to move away from the shock. In this scenario proton never enters the downstream region. 
\item A proton that is transmitted to the downstream region does not return to the upstream region but continues its spiral motion in downstream region. These two possibilities are  actually the same as in the case of a longitudinal shock.
\item The last possibility is somewhat more complicated. A proton can be either reflected to the upstream region or transmitted to the downstream region. Then, due to its helical motion, it can enter the shock region again and repeat this a couple of times (Figure~\ref{fig34}). This scenario is of course only possible within the Larmor from the shock and since we are only only focusing on a near-shock region, we are not interested in the effects occurring in sheath region far away (compared to shock thickness) from the shock.
\end{enumerate}

Whatever the final outcome is, whether the proton ends up in upstream or downstream region, it is important that once the proton is in the downstream region it can be detected as transmitted. Since some protons will come deeper into the downstream region than others, we will get a smaller number of protons farther away from the shock. So the Forbush decrease amplitude is the smallest near the shock and it is given by the number of particles that never enter the downstream region. The maximum of the amplitude is given by the particles that enter the downstream region, and do not return to the upstream region.

\section{Analysis of the Parameter Space}
Now we examine how a shock with given parameters (shock thickness $ d $, magnetic field in the upstream region $B_{1}$, inclination angle $\theta $ and field ratio $B_{2}/B_{1}$) affects proton behavior. We also vary proton energy, \textit{i.e.} proton speed. As our reference point, we take a proton of 1~ GeV energy, $B_{1}=5$~nT and $B_{2}/B_{1}=4$ which is the upper limit for perpendicular shock. Typical span of values for shock thickness is between $40 $~km and $120~000$~km \citep{pinter80}. We take for the shock thickness to be $d=1\ 000$~km which is approximately the value given in Table~3 in \citeauthor{pinter80},  (\citeyear{pinter80}). Larger thickness gives particle more time to reflect (\textit{i.e.} more time for $v_{x}$ to drop to zero) and also results in larger number of reflected particles. This is different from the magnetic mirror case where mirror condition does not depend on the length of trap. However, we have already emphasized that we are not using the magnetic mirror approach because the shock thickness is smaller than the gyro-radius of the proton.

\begin{figure*}[htbp]
\centering
\subfloat[]{\includegraphics[width=0.5\textwidth]{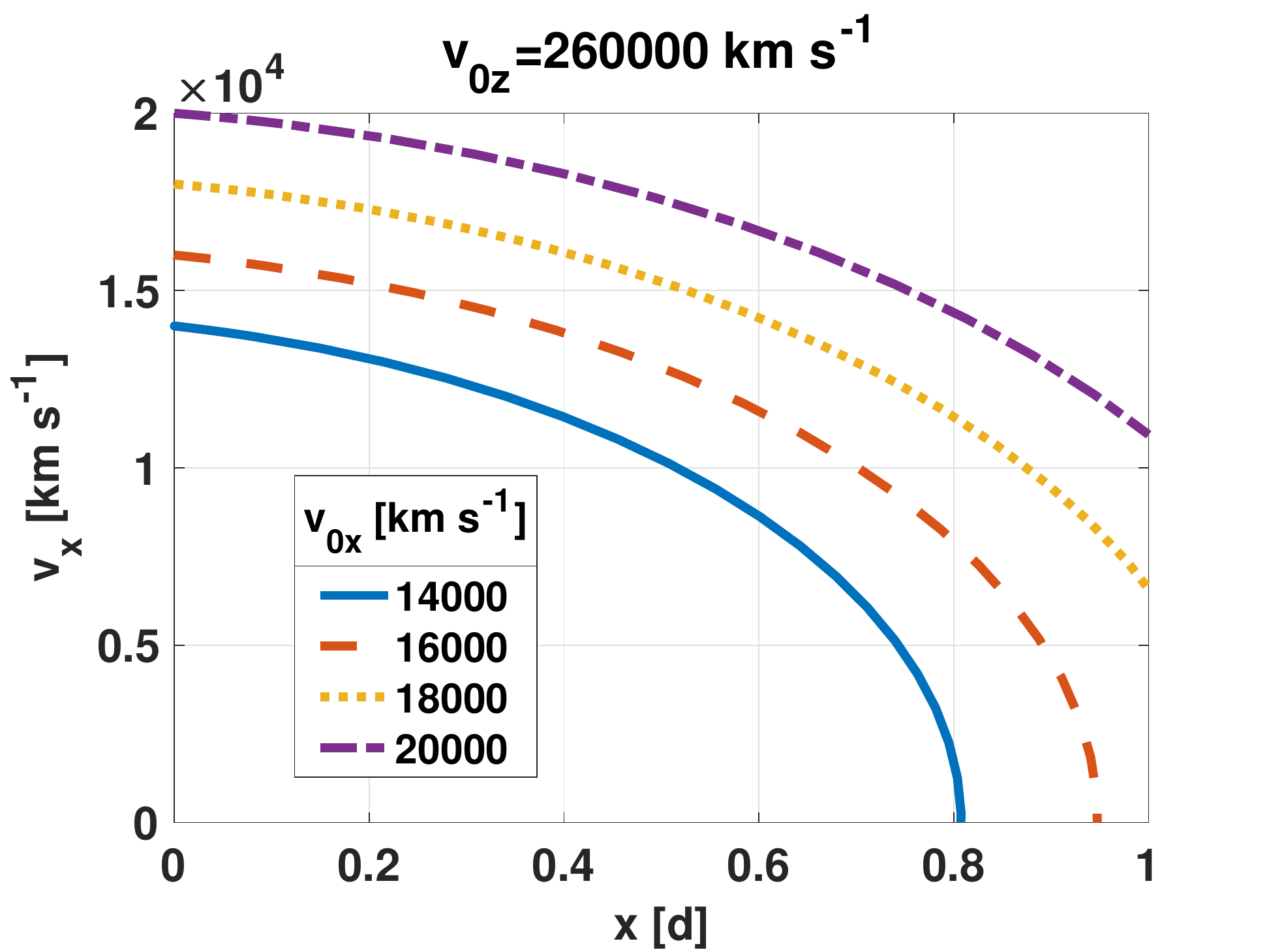}\label{fig41a}}
\subfloat[]{\includegraphics[width=0.5\textwidth]{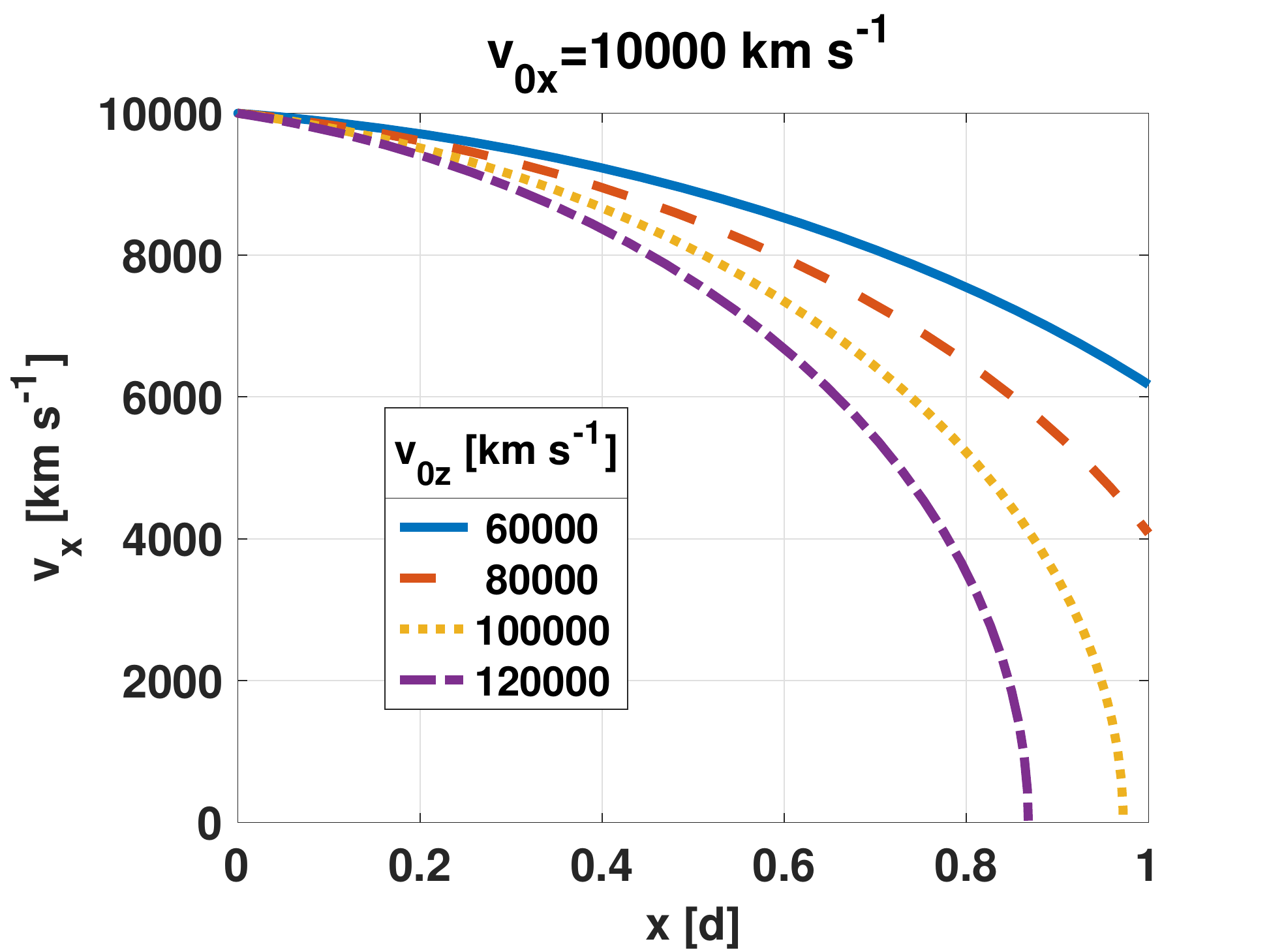}\label{fig42b}}
\caption{Change of $x$-component of proton speed inside an oblique shock ($B_{1}=5.0 $~nT, $B_{2}/B_{1}=4$ and $\theta=45^{\circ} $). Initial conditions are varied for a proton of 1~ GeV energy and 1~000~km shock thickness; a)~shows the change of $v_{x}$ with distance from the shock border for different initial values of $v_{x}$ and constant initial value of $v_{z}$, namely $v_{0z}=260\ 000 $~km~s$^{-1}=0.867 c$; b)~shows the change of $v_{x}$ with distance from the shock for different initial values of $v_{z}$ and constant initial value of $v_{x}$, namely $v_{0x}=10\ 000$~km~s$^{-1}=0.033c$. When $v_{x}$ becomes zero inside the shock, \textit{i.e.} for $x<1$, it will continue to decrease and become negative which means that the proton is reflected.}
\label{fig41}
\end{figure*}

Figure~\ref{fig41} shows the change of $v_{x}$ component of proton speed inside the shock for different initial conditions. In Figure~\ref{fig41}a we vary the initial value of  $v_{x}$ while the initial value of $v_{z}$ remains the same. For a 1~GeV proton and initial value $ v_{0z}=260\ 000$~km~s$^{-1}=0.867 c$, $v_{0x} $ has to be smaller than approximately $17\ 000$~km~s$^{-1}=0.057c$ (which is more than 10 times smaller than $v_{0z} $) for the proton to be reflected. On the other hand, if we keep a constant initial value $ v_{0x}=10\ 000$~km~s$^{-1}=0.033c$ , and vary the initial value of $ v_{z} $ (Figure~\ref{fig41}b), we see that the initial value $ v_{0z} $ has to be higher than $100\ 000$~km~s$^{-1}=0.333 c$  for a proton to be reflected. For a proton to be reflected no matter what the initial value $v_{0x}$ was, $ v_{0z}$ limiting value should be $261\ 918$~km~s$^{-1}=0.874 c$, which means $99.80 \% $ of total proton speed for 1~GeV proton.

\begin{figure*}[h!]
\centering
\subfloat[]{\includegraphics[width=0.5\textwidth]{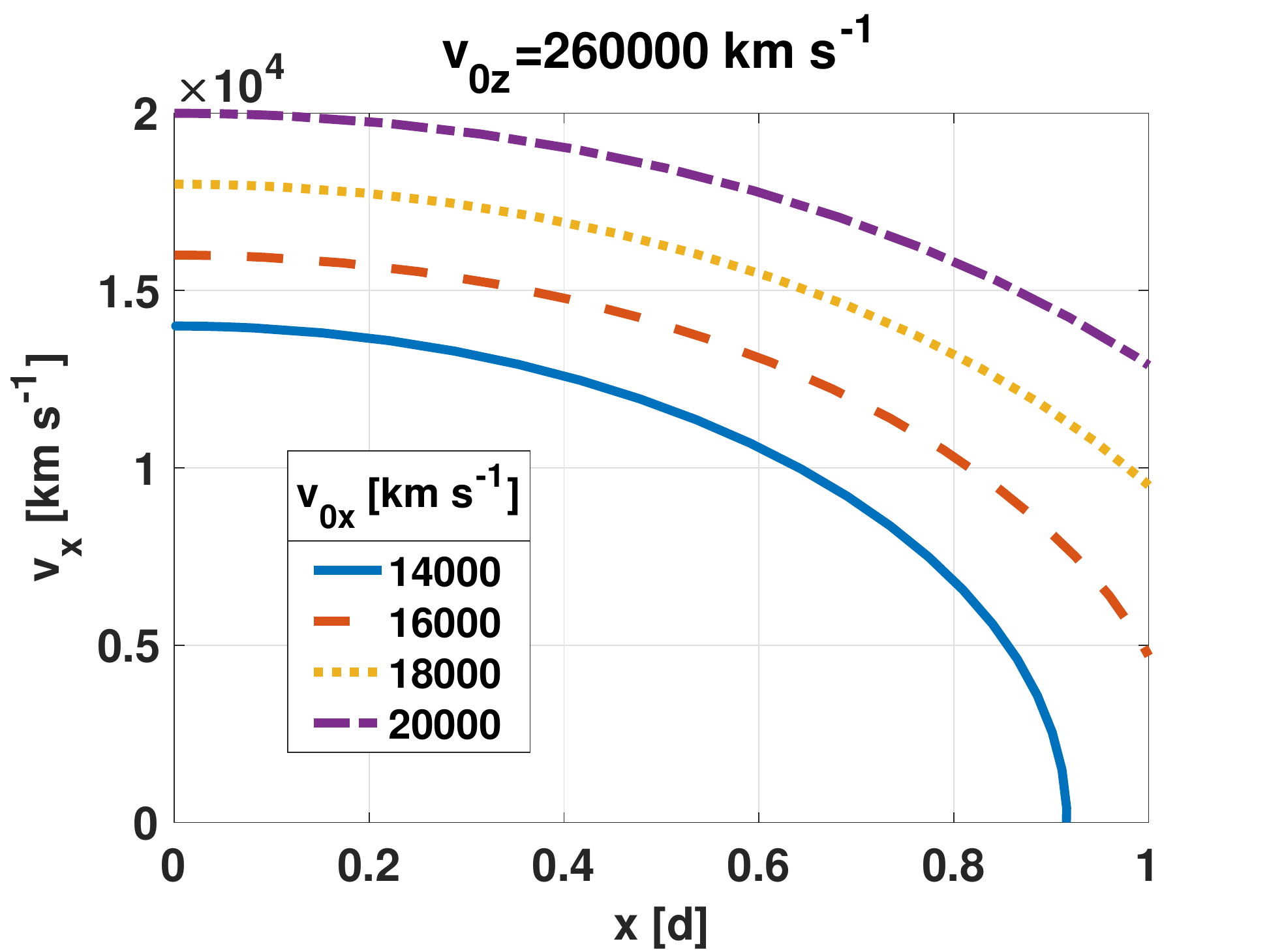}\label{fig42a}}
\subfloat[]{\includegraphics[width=0.5\textwidth]{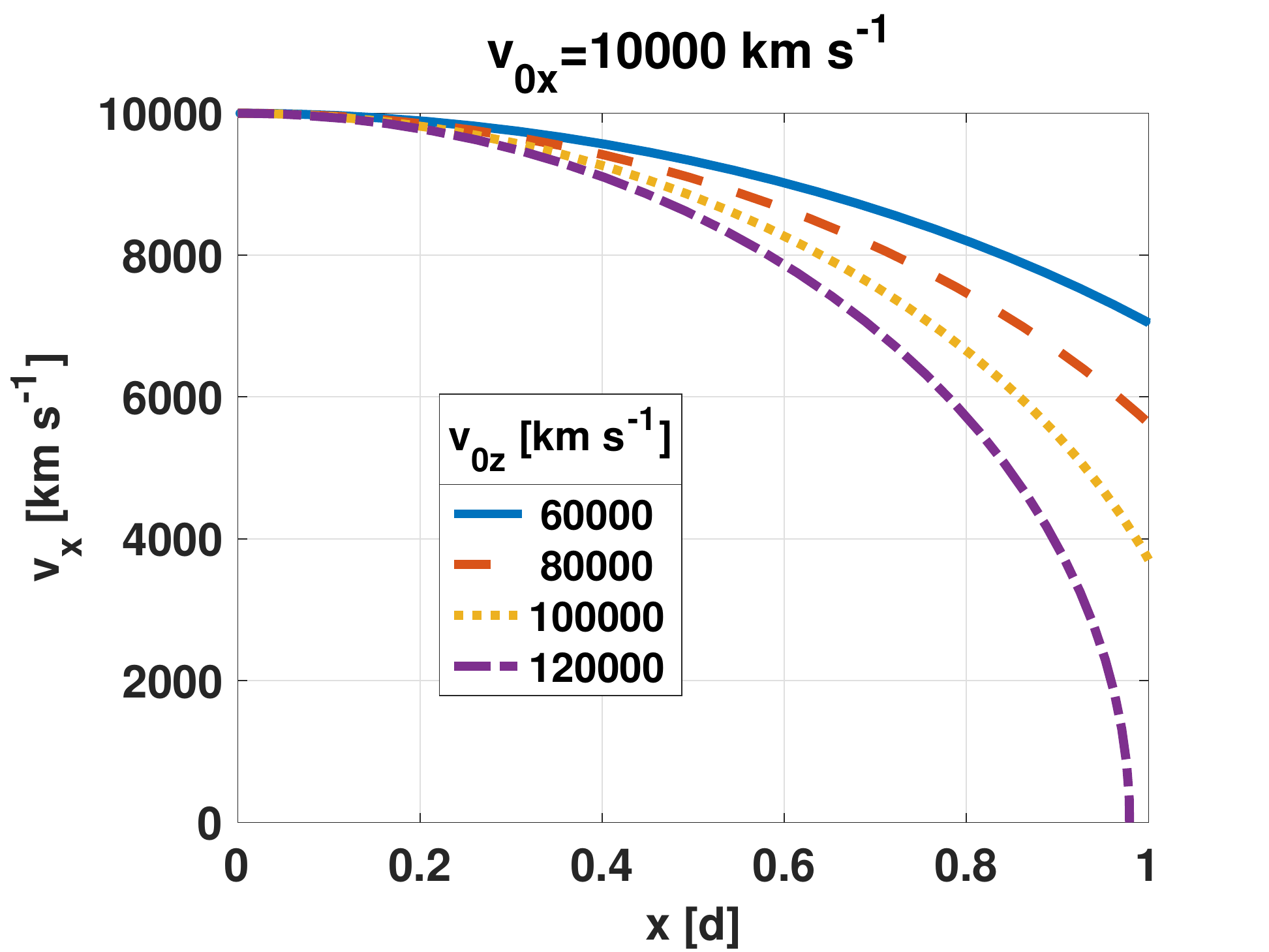}\label{fig42b}}\\
\caption{Change of $x$-component of proton speed inside the longitudinal shock for a proton of 1~GeV energy and $1\ 000$~km shock thickness ($B_{1}=5$~nT and $B_{2}/B_{1}=4$).}
\label{fig42}
\end{figure*}

Figure~\ref{fig42} shows that the situation for a longitudinal shock is qualitatively the same, and even quantitatively very similar. The total field is the same as in previous example (5~nT), but the shock inclination angle and therefore field components are different. As it can be seen in graphs, in fields with weaker \textit{y}-component (\textit{i.e.} longitudinal shocks) $v_{x}$ will drop to zero for somewhat larger values of $v_{0z}$. This is in agreement with Equation~(\ref{ax}) where stronger $B_{1y}$ results in stronger deceleration $a_{x}$. Similar conclusion could be made by ``putting'' the particle at a perpendicular shock but, since in that case the particle  should gyrate parallel to the shock, we omit that analysis here.

In the next step we establish the critical values of $v_{0x}$ and $v_{0z}$ for which the proton is reflected the moment before it would pass the shock (at $x\approx d$). For all the $v_{0x}$ smaller, or $v_{0z}$ larger than those critical values, proton would be reflected inside the shock. This gives us a range of values over which can be integrated to obtain a percentage of protons that are reflected. Since the total proton speed is constant, the value of $v_{0y}$ is determined by chosen $v_{0x}$ and $v_{0z}$. However, the sign of $v_{0z}$ plays an important role, especially at small $v_{0x}$ and $v_{0z}$ when $v_{0y}$ is large. Negative and large $v_{0y}$ gives large and positive $a_{z}$ which increases rapidly $v_{0z}$ which in return decreases $v_{0x}$. This can be seen in Table~\ref{tab1} where for a given $v_{0x}$, the critical values of $v_{0z}$ are given for positive and negative values of $v_{0y}$. If $v_{0x}$ is higher than approximately $16\ 800$~km~s$^{-1}=0.056c$ the proton will not be reflected no matter what the speed $v_{0z}$ is. For $v_{0x} $ close to this value, critical values of $v_{0z}$ are roughly the same for positive and negative $v_{0y}$. However, for smaller $v_{0x}$ the critical values of $v_{0z}$ differ greatly for positive and negative $v_{0y}$.

\begin{table}[h!]
\begin{tabular}{lccccccc}
\hline
$v_{0x}$                   & 1 000   &    2 000  & 4 000    & 8 000 & 12 000  & 16 000 & 16 800 \\ \hline
$v_{0z} (v_{0y}>0)$   & 19 100 &  22 400  & 24 200  &    63 000   & 135 400  &  237 900 &  261 900 \\ 
$v_{0z} (v_{0y}<0)$   &     0     &        0    &   6 900  &  55 000     &  130 800  &  236 000 &  261 500   \\ \hline
\end{tabular}
\caption{Critical values of $v_{0z}$ for different $v_{0x}$ for positive (second row) and negative (third row) $v_{0y}$. The values are given in km~s$^{-1}$. Field parameters are  $B_{1}=5.0 $~nT, $B_{2}/B_{1}=4$, $\theta=45^{\circ}$ and $d=1\ 000$~km.}
\label{tab1}
\end{table}

For 1~GeV protons, $v_{0x}=16\ 800$~km~s$^{-1}=0.056c$ is the maximum value of \textit{x}-component of initial speed which allows the proton to be reflected. In this case, $v_{0z}$ must be at least 261~900~km~s$^{-1}=0.874c$. If we increase proton energy to 10~GeV, we get the maximum value of $v_{0x}$ to be 7~550~km~s$^{-1}=0.025c$. For even higher energies of, for example, 100~GeV the maximum value of $v_{0x}$ is even smaller, $2\ 480$~km~s$^{-1}=0.008c$. At the same time the initial value of $v_{0z}$ relative to $v_{0}$ stays almost the same (Table~\ref{tab2}). This indicates that there will be a smaller percentage of reflected particles at higher energies, \textit{i.e.} Forbush decrease will be stronger for lower energy particles in agreement with observational studies showing rigidity dependence of FD magnitude (see \textit{e.g.} \citeauthor{lockwood71}, \citeyear{lockwood71}; \citeauthor{cane00}, \citeyear{cane00} and references therein).

\begin{table}[]
\begin{tabular}{cccc}
\hline
E {[}GeV{]} & $v_{0x}$ {[}km s$^{-1}${]} & $v_{0z}$ {[}km s$^{-1}${]} & $ v_{0z}/v_{0}$ \\ \hline
1           & 16 800         & 261 800        & 0.9974 \\ 
10          & 7 550          & 298 400        & 0.9983 \\ 
100         & 2 490          & 299 500        & 0.9983 \\ \hline
\end{tabular}
\caption{Critical values of $v_{0z}$ for different $v_{0x}$ and different proton energies. Field parameters are  $B_{1}=5.0 $~nT, $B_{2}/B_{1}=4$, $\theta=45^{\circ}$ and $d=1\ 000$~km.}
\label{tab2}
\end{table}

Lowering the ratio $B_{2}/B_{1}$ results in a weaker magnetic field behind the shock, $B_{2}$, but also in weaker magnetic field inside the shock which should in turn cause smaller Forbush decrease, qualitatively in agreement with observational studies showing that FD magnitude is larger for larger fields (\textit{e.g.} \citeauthor{belov01}, \citeyear{belov01}; \citeauthor{dumbovic11},  \citeyear{dumbovic11}, \citeyear{dumbovic12}; \citeauthor{richardson11}, \citeyear{richardson11}). If we look at the largest possible initial values of $v_{x}$ (Table~\ref{tab3}) we see that they are indeed lower than in previous cases while the corresponding $v_{0z}$ remains almost the same.

\begin{table}[]
\begin{tabular}{cccc}
\hline
E {[}GeV{]} & $v_{0x}$ {[}km s$^{-1}${]} & $v_{0z}$ {[}km s$^{-1}${]} & $ v_{0z}/v_{0}$ \\ \hline
1           & 12 500         & 261 300        & 0.9956\\ 
10          & 5 600          & 298 700        & 0.9993 \\ 
100         & 1 855          & 299 500        & 0.9983 \\ \hline
\end{tabular}
\caption{Critical values of $v_{0z}$ for different $v_{0x}$ and different proton energies. Field parameters are  $B_{1}=5.0 $~nT, $B_{2}/B_{1}=2$, $\theta=45^{\circ}$ and $d=1\ 000$~km.}
\label{tab3}
\end{table}

Similar effect, but stronger, arises when we decrease the value of the magnetic field in the upstream region, \textit{i.e.} $B_{1}$. If we set $B_{1}=2.5 $~nT and $B_{2}/B_{1}=4$ which results with the same $B_{2}$ as in the previous case, we get even smaller maximum values of $v_{0x}$ (see Table~\ref{tab4}). In this case the field component $B_{1y}$ is smaller so the \textit{x}-component of acceleration in Equation~(\ref{ax}) is smaller and hence it takes longer time for $v_{x}$ to become zero. 

\begin{table}[]
\begin{tabular}{cccc}
\hline
E {[}GeV{]} & $v_{0x}$ {[}km s$^{-1}${]} & $v_{0z}$ {[}km s$^{-1}${]} & $ v_{0z}/v_{0}$ \\ \hline
1           & 11 800         & 261 900        & 0.9979\\ 
10          & 5 340          & 298 600        & 0.9990 \\ 
100         & 1 760          & 299 300        & 0.9977 \\ \hline
\end{tabular}
\caption{Critical values of $v_{0z}$ for different $v_{0x}$ and different proton energies. Field parameters are  $B_{1}=2.5 $~nT, $B_{2}/B_{1}=4$, $\theta=45^{\circ}$ and $d=1\ 000$~km.}
\label{tab4}
\end{table}

Finally, Table~\ref{tab5} proves the statement we made at the beginning: for thicker shocks, \textit{i.e.} for larger values of $d$ in Equation~(\ref{ax}), larger values of $v_{0x}$ are allowed because even though the deceleration $a_{x}$ is smaller, there is more time for a particle to get reflected.  

\begin{table}[]
\begin{tabular}{cccc}
\hline
E {[}GeV{]} & $v_{0x}$ {[}km s$^{-1}${]} & $v_{0z}$ {[}km s$^{-1}${]} & $ v_{0z}/v_{0}$ \\ \hline
1           & 52 800         & 256 600        & 0.9976\\ 
10          & 23 800          & 296 400        & 0.9916 \\ 
100         & 780          & 294 000        & 0.9800 \\ \hline
\end{tabular}
\caption{Critical values of $v_{0z}$ for different $v_{0x}$ and different proton energies. Field parameters are  $B_{1}=5.0 $~nT, $B_{2}/B_{1}=4$, $\theta=45^{\circ}$ and $d=10\ 000$~km.}
\label{tab5}
\end{table}

\section{Conclusion}
We presented a model for magnetic field linearly changing inside an oblique 2-dimensional MHD fast-mode shock and used that model to determine whether a proton of certain energy will be reflected inside the shock. We vary initial conditions and solve numerically the set of three differential equations (one for each speed component). The goal is to find such initial conditions that the proton is reflected just at the farther end of the shock. The corresponding initial values of $v_{x}$ and $v_{z}$ are called critical values. All $v_{0x}$ smaller than the critical value of $v_{0x}$ and all $v_{0z}$ larger than the critical value of $v_{0z}$ would also result in reflected protons. Obtaining this critical values is an important step in calculating the Forbush decrease amplitude for protons of certain energy. Since we are adjusting initial values, they are only approximate. Their accuracy is within 100 km s$^{-1}$ which is less than 0.04\% of the proton speed so the uncertainty is too small to affect the final outcome. Results demonstrate that protons with higher energies are less likely to be reflected. Also, thicker shocks and shocks with larger $B_{1y}$ field component reflect more protons than those with smaller. Our next step is to use the obtained results and calculate the Forbush decrease amplitude for different field configurations and particle energies.

%%%%%%%%%%%%%%%%%%%%%%%%%%%%%%%%%%%%%%%%%%%%%%%%%%%%%%%%%%%%%%%%%%%%%%%%%%%
%% Appendix
%
% \appendix   
%%%%%%%%%%%%%%%%%%%%%%%%%%%%%%%%%%%%%%%%%%%%%%%%%%%%%%%%%%%%%%%%%%%%%%%%%%%
%% Acknowledgements
%
 \begin{acks}
BV and MD acknowledge a support by the Croatian Science Foundation under the project 7549 ``Millimeter and submillimeter observations of the solar chromosphere with ALMA". The research leading to these results has received funding from the European Union’s Horizon 2020 research and innovation programme under the Marie Sklodowska-Curie grant agreement No 745782 (ForbMod).
 \end{acks} 

\paragraph*{Disclosure of Potential Conflicts of Interest}  The authors declare that they have no conflicts of interest.

%%% %%%%%%%%%%%%%%%%%%%%%%%%%%%%%%%%%%%%%%%%%%%%%%%%%%%%%%%%%%%
%% Bibliography
%
% Using BibTeX
%
\bibliographystyle{spr-mp-sola}
\bibliography{bibliography}

\end{article} 
\end{document}